\documentstyle[preprint,prl,aps]{revtex} 

\newcommand{\Tr}{{\rm Tr}\,}

\newcommand{\CN}{{\cal N}}

\newcommand{\CF}{{\cal F}}
\newcommand{\CZ}{{\cal Z}}

\newcommand{\half}{\frac{1}{2}}
\font \barfont = bbm10
\newcommand{\R}{\mbox{\bf\barfont R}}
 
\begin{document}
\draft
\title{\begin{flushright}
{\small\hfill AEI-2000-037\\
\hfill hep-th/0006234}\\
\end{flushright}
Bulk Witten Indices and the Number of Normalizable Ground States
in Supersymmetric Quantum Mechanics of Orthogonal, Symplectic and
Exceptional Groups}
\author{Matthias Staudacher \footnote{matthias@aei-potsdam.mpg.de }
}
\address{
Albert-Einstein-Institut, Max-Planck-Institut f\"{u}r
Gravitationsphysik\\ Am M\"uhlenberg 1\\  D-14476 Golm, Germany }
\maketitle
\begin{abstract}

This note addresses the question of the number of normalizable vacuum states
in supersymmetric quantum mechanics with sixteen supercharges and
arbitrary semi-simple compact gauge group, up to rank three.
After evaluating certain contour integrals obtained by appropriately
adapting BRST deformation techniques we propose
novel rational values for the bulk indices. 
Our results demonstrate that an asymptotic method for obtaining 
the boundary contribution to the index, originally due to Green and Gutperle,
fails for groups other than SU$(N)$. We then obtain
likely values for the number of ground states of these systems.
In the case of orthogonal and symplectic groups our finding is
consistent with recent conjectures of Kac and Smilga,
but appears to contradict their result in the case
of the exceptional group $G_2$.

\end{abstract}
\vspace{5.5cm}
\newpage
\narrowtext

Supersymmetric Yang-Mills theories dimensionally reduced to zero spacial
dimensions were initially considered as interesting examples for
susy quantum mechanics \cite{ch},\cite{smilga}. The Hamiltonian
of these systems reads
\begin{equation}
H={1 \over 2 g^2} \Tr \Bigg( P_i P_i -\half 
[X_i,X_j] [X_i,X_j] -
\Psi_{\alpha} [ \Gamma_{\alpha \beta}^i X_i,\Psi_{\beta}] \Bigg).
\label{hamilton}
\end{equation}
where the bosonic ($X_i$) and fermionic ($\Psi_i$) 
degrees of freedom take values in the
Lie algebra of the compact gauge group. 
Due to the representation theory of the gamma matrices 
$\Gamma_{\alpha \beta}^i$ this quantum mechanics only exists
if the number of supercharges is $\CN=$2,4,8 or 16, 
corresponding to the dimensional
reduction of $\CN=1$ supersymmetric gauge field theory in $D=d+1$ dimensions,
where $d=2,3,5,9$, respectively (i.e.~$i=1, \ldots, d$).

The $d=9$ system gained relevance following work of de Wit, Hoppe
and Nicolai \cite{dhn}, who argued that the light cone quantization of
11-dimensional supermembranes could be described by the model
in eq.(\ref{hamilton}) with gauge group SU$(N)$ in the limit 
$N \rightarrow \infty$. This interpretation rendered more urgent the
question about the Hamiltonian's spectrum. It was quickly understood
that the latter is continuous and, in fact, that there are non-localized
states for any positive energy eigenvalue \cite{smilga},\cite{dln}.
As far as the supermembrane is concerned, this was initially
considered to be an unphysical feature. More recently, the model
was ``resuscitated'' as a proposed formulation of M-theory \cite{bfss},
albeit for a special background, and, interestingly,
the continuous spectrum was turned into a virtue. A crucial issue,
required in all known applications of the $d=9$ system eq.(\ref{hamilton}),
is whether there also exists a {\it normalizable} zero energy vacuum state
(which may loosely be called the ``graviton multiplet''). On the other
hand, from various points of view, one does not expect such a state in
the cases $d=2,3,5$. A rigorous proof for $d=3,5,9$ 
of this has so far only been proposed
for SU$(2)$ \cite{sestern} (see also \cite{yi}). 
For alternative insights into
this question see \cite{hoppe1},\cite{hoppe2},\cite{porrati}.

For the group $G=$SU$(N)$, 
the proof has not yet been fully completed, but a strategy generalizing 
the method of \cite{yi},\cite{sestern}, as well as some important
partial results, exists. In fact, the procedure immediately applies to more
general compact gauge groups $G$. We will exclude from now
on discussion of the case $d=2$, where the methods below appear to
fail. The idea is to compute the Witten index
ind$^{D=d+1}(G)$ of the quantum mechanics eq.(\ref{hamilton}):
\begin{equation}
{\rm ind}^D(G)={\rm lim}_{\beta \rightarrow \infty} \Tr (-1)^F e^{-\beta H}=
n^0_B-n^0_F
\label{ind}
\end{equation}
giving the number of $n^0_B$ bosonic minus $n^0_F$ fermionic 
zero energy states. Clearly ind$^D(G)$ gives a lower bound on the number
of vacuum states. This lower bound has recently been argued to be saturated
for the systems we are studying \cite{sestern2}. 
In light of the above we would expect
${\rm ind}^{D=10}($SU$(N))=1$ and ${\rm ind}^{D=4,6}($SU$(N))=0$. 
Using heat kernel methods, it is technically much easier to
calculate the so-called bulk index 
\begin{equation}
{\rm ind}_0^D(G)={\rm lim}_{\beta \rightarrow 0} \Tr (-1)^F e^{-\beta H}
\label{bind1}
\end{equation}
which may be related to a finite susy {\it Yang-Mills integral} 
$\CZ^{\CN}_{D,G}$:
\begin{equation}
{\rm ind}_0^D(G)=
\frac{1}{\CF_G} \CZ^{\CN}_{D,G}.
\label{bind2}
\end{equation}
This is an ordinary (as opposed to functional) multiple integral given by
\begin{equation}
\CZ^{\CN}_{D,G}:=\int \prod_{A=1}^{{\rm dim}(G)} 
\Bigg( \prod_{\mu=1}^{D} \frac{d X_{\mu}^{A}}{\sqrt{2 \pi}} \Bigg) 
\Bigg( \prod_{\alpha=1}^{\CN} d\Psi_{\alpha}^{A} \Bigg)
\exp \bigg[  \frac{1}{4 g^2} \Tr 
[X_\mu,X_\nu] [X_\mu,X_\nu] + \frac{1}{2 g^2} 
\Tr \Psi_{\alpha} [ \Gamma_{\alpha \beta}^{\mu} X_{\mu},\Psi_{\beta}]
\bigg],
\label{susyint}
\end{equation}
where dim$(G)$ is the dimension of the Lie group and the $D=d+1$ bosonic
matrices $X_{\mu}=X_{\mu}^A T^A$ and the $\CN$ fermionic matrices
$\Psi_{\alpha}=\Psi_{\alpha}^A T_A$ 
are anti-hermitean and take values in the
fundamental representation of the Lie algebra Lie$(G)$,
whose generators we denote by $T^A$. $g^2$ is fixed according to
the normalization $\Tr T^A T^B=-g^2 \delta^{AB}$.
The constant $\CF_G$ in eq.(\ref{bind2}) is 
essentially the volume of the true gauge group, which turns
out to be the quotient group $G/Z_G$, with $Z_G$ the center group of $G$.
For more details see \cite{ks3}.

The integral eq.(\ref{susyint}) is still very complicated and has so far
only been directly analytically calculated for $G=$SU(2) 
\cite{smilga},\cite{yi},\cite{sestern}. It has however been indirectly
calculated for SU$(N)$ by supersymmetric BRST deformation techniques
in \cite{mns}. The derivation involved some assumptions and
unproven steps, but the result has been confirmed for various values of
$N$ and $D$ in numerical (Monte Carlo) studies \cite{us}.
The result may be summarized in the following table.
\begin{center}
Table 1: $D=4$, $D=6$ and $D=10$ bulk index ind$_0^D$
and (for $N>2$, conjectured) total Witten index ind$^D$ for the 
special unitary groups of arbitrary rank. \\
\vspace{0.5cm}
\begin{tabular}{||c|| c | c | c | c | c | c ||} \hline
Group  & rank  & ind$_0^{D=4,6}$ & ind$^{D=4,6}$ & ind$_0^{D=10}$ &
ind$^{D=10}$ \\ \hline  \hline  
SU(N)  & $N-1$ & $1/N^2$  & 0  & $\sum_{m|N} 1/m^2$ & 1 \\ \hline
\end{tabular} 
\end{center}
We see that ind$^D \neq$ ind$_0^D$ contrary to what one might have suspected:
The continous spectrum renders $\Tr (-1)^F e^{-\beta H}$ $\beta$-dependent,
and the full Witten index ind$^D$ is given by
\begin{equation}
{\rm ind}^D(G)={\rm ind}^D_0(G)+{\rm ind}^D_1(G).
\label{full}
\end{equation}
where ind$_1^D$ is a ``boundary term'' that is picked up when going from
eq.(\ref{ind}) to eq.(\ref{bind1}). Its value has been rigorously
evaluated for SU(2) in \cite{sestern}. For general special unitary
groups, it has been argued by Green and Gutperle \cite{greengut}
that the boundary
term may be deduced by considering a {\it free} effective
Hamiltonian for the diagonal (Cartan) degrees of freedom of the
$d$ Lie-algebra valued fields $X_i$, with the discrete Weyl symmetry
of $G$ imposed as a constraint on the wave functions. This led to
the evaluation of a simple Gaussian integral and in consequence to
\begin{equation}
{\rm ind}_1^{D=4,6}({\rm SU}(N))=-\frac{1}{N^2} 
\qquad {\rm and} \qquad 
{\rm ind}_1^{D=10}({\rm SU}(N))=-\sum_{m|N \atop m>1} \frac{1}{m^2}.
\label{boundary}
\end{equation} 
Incidentally, the $D=10$
expression was proposed before the evaluation of the bulk indices
\cite{greengut}.
It seems therefore very likely that the numbers of table 1 are
indeed correct for all $N$. In order to further test these ideas,
it is clearly of interest to apply them to various other gauge
groups. As we shall see, most of the features just discussed
become more intricate.

The zero-energy bound state problem for the Hamiltonian eq.(\ref{hamilton})
for general semi-simple gauge groups was recently considered 
in detail by
Kac and Smilga \cite{kasm}. Generalizing the mass deformation
method of \cite{porrati}, these authors made the very interesting claim 
that the Yang-Mills
quantum mechanics can lead for $D=10$ to {\it more} than one
vacuum state. Assuming that all large mass bound states remain
normalizable as the mass is tuned to zero -- this is the potential
weak point of the method -- they found
\[
{\rm ind}^{D=10}({\rm SO}(N))={\rm  number~of~partitions~of}~N 
~{\rm into~distinct~odd~parts}
\]
\[
\;\;\; {\rm ind}^{D=10}({\rm Sp}(2N))={\rm number~of~partitions~of}~2 N
~{\rm into~distinct~even~parts}
\]
\[
{\rm ind}^{D=10}({\rm G}_2)=2 \quad {\rm ind}^{D=10}({\rm F}_4)=4
\]
\begin{equation}
{\rm ind}^{D=10}({\rm E}_6)=3 \quad {\rm ind}^{D=10}({\rm E}_7)=6 \quad 
{\rm ind}^{D=10}({\rm E}_{8})=11
\label{states}
\end{equation}
In the case of the SO$(N)$ and Sp$(2N)$, 
Hanany et.al.~presented independent arguments in favor of these 
multiplicities by considering a ``physical'' application of the
$d=9$ Yang-Mills quantum mechanics to orientifold 
points in M theory \cite{hanany}.

Clearly it is worthwhile to apply the index computations of the last
section to the new classes of groups. In fact, Kac and Smilga \cite{kasm}
generalized
the Green-Gutperle \cite{greengut} method based on a free effective
Hamiltonian, and, after imposing the discrete Weyl symmetry and 
performing the appropriate Gaussain integration,
proposed for $D=4,6$ the explicit formula
\begin{equation}
{\rm ind}_1^{D=4,6}(G)=^{?}-\frac{1}{|W_G|} {\sum_{w \in W_G}}' 
~\frac{1}{\det (1-w)}
\label{weylsum}
\end{equation}
where the sum $\sum^{'}$ extends over all elements $w$ of the
Weyl group $W_G$ of $G$ such that det$(1-w) \neq 0$.
They also derived a slightly more involved $D=10$ expression.
For $G=$SU$(N)$ one quickly rederives eq.(\ref{boundary}) from 
these expressions.

One should next compute the bulk indices ind$_0^D(G)$. It is 
straightforward, in principle, to extend the deformation 
method of Moore et.al.~\cite{mns} to the new groups. This was done
in \cite{ks3} for $D=4$ and rank up to three. We carefully evaluated the
corresponding (see eq.(\ref{bind2}))
Yang-Mills integrals eq.(\ref{susyint}) by Monte Carlo
and verified that the BRST deformation method indeed applies to
the new groups as well. Surprisingly, the result did {\it not}
match the expression eq.(\ref{weylsum}) (remember that we 
expect ind$_0^{D=4}+$ind$_1^{D=4}=0$). {\it This indicates failure
of the description based on an effective free Hamiltonian.}    
 
In the present note we will apply the BRST deformation method \cite{mns}
to also work out the bulk indices of the $D=10$ model. This method
consists in adding cubic and quadratic terms to the action which break
all but one of the supersymmetries. The remaining symmetry still assures
that the partition function remains unchanged. By taking limits the
non-linearities of the original action are simply dropped, and all
integrations can be performed. For $D=10$ one finds the $r$-fold integral
\[
{\rm ind}_0^{D=10}(G)=\frac{|Z_G|}{|W_G|}~
C^r 
~\oint \prod_{k=1}^r \frac{d x_k}{2 \pi i}
\Delta_G(0,x) {\Delta_G(E_1+E_2,x) \Delta_G(E_1+E_3,x) \Delta_G(E_2+E_3,x) 
\over
\Delta_G(E_1,x) \Delta_G(E_2,x) \Delta_G(E_3,x) \Delta_G(E_4,x) }
\]
\begin{equation}
C= {(E_1+E_2)(E_1+E_3)(E_2+E_3) \over E_1+E_2+E_3+E_4} 
\label{contour}
\end{equation}
where $|Z_G|$ and $|W_G|$ are the orders of, respectively, the
center group $Z_G$ and Weyl group $W_G$ of $G$, and $r$ is the rank
of $G$.
For the various groups one has (see \cite{ks3} for details):
\[
\Delta_{{\rm SO}(2N+1)}(E,x)=
\prod_{i<j}^N
\Big[(x_i-x_j)^2-E^2\Big] \Big[(x_i+x_j)^2-E^2\Big]
\prod_{i=1}^N \Big[x_i^2-E^2 \Big]
\]
\[
\Delta_{{\rm Sp}(2N)}(E,x)=
\prod_{i<j}^N
\Big[(x_i-x_j)^2-E^2\Big] \Big[(x_i+x_j)^2-E^2\Big]
\prod_{i=1}^N \Big[x_i^2-(\frac{E}{2})^2 \Big]
\]
\[
\Delta_{{\rm SO}(2N)}(E,x)=
\prod_{i<j}^N
\Big[(x_i-x_j)^2-E^2\Big] \Big[(x_i+x_j)^2-E^2\Big]
\]
\[
\Delta_{{\rm G}_2}(E,x)=
\Big[(x_1-x_2)^2-E^2 \Big]
\Big[ (x_1+x_2)^2-E^2 \Big]  \Big[ x_1^2-E^2 \Big] \Big[ x_2^2-E^2 \Big]
\Big[(2 x_1 +x_2)^2-E^2 \Big] \Big[ (x_1+2x_2)^2-E^2 \Big]
\]
The integrals eq.(\ref{contour}
are divergent as ordinary integrals over $\R^r$;
if however one (a) interprets them as contour integrals, and (b)
gives a small imaginary part to the real parameters 
$E_j \rightarrow E_j + i \epsilon$ they converge. The resulting number
is, after imposing 
${\rm Re}(E_4)=-{\rm Re}(E_1)-{\rm Re}(E_2)-{\rm Re}(E_3)$,  
independent\footnote{This is actually not entirely true. If the $E_j$ are
rational numbers there are numerous ``special'' configurations where simple
poles merge to higher order poles in the denominator of the integrand and
the integral gives a wrong value. The correct bulk index is only obtained
for a ``generic'' pole configuration.}
of the parameters $E_j$. As we already remarked, the above presciptions
would have to be derived from first principles if one wants to
render the BRST deformation method entirely rigorous, but we have good
evidence \cite{us},\cite{ks3} that it really works for $D=4,6,10$. 

It would be interesting to evaluate the $D=4,6,10$ 
contour integrals for general SO$(2N+1)$, Sp$(2N)$ and SO$(2N)$,
as has been done (at least for $D=4,6$) for SU$(N)$ in \cite{mns}.
In particular, it would be exciting to find the correct replacement for
eq.(\ref{weylsum}).
This has to date proved too difficult; what can be done is an evaluation
for small rank. For $D=10$ even this is rather involved, and we needed
to apply symbolic manipulation programs to locate and evaluate the
huge number of residues for the contour integrals eq.(\ref{contour}).
The results, for rank up to three, are shown in the table 2 below.

We would next like to be able to find the total index ind$^{D=10}(G)$
in order to verify the conjectures eq.(\ref{states}). 
As we have demonstrated already for $D=4,6$, and may be verified
for $D=10$ as well by comparing the numbers in table 2 with those
of \cite{kasm}, we are currently lacking a reliable method
for determining the boundary index ind$_1^{D=10}(G)$.
Given only the bulk index ind$_0^{D=10}(G)$, can we still make a statement
about the number of vacuum states? In all known cases the boundary
index is a negative rational number between zero and one. Assuming
this to be true in general, we are able to predict the Witten index,
which has to be an integer. This prediction is shown in the rightmost
column of table 2. Comparing to eq.(\ref{states}), we find agreement
for the orthogonal and symplectic groups of rank $\leq3$ (in particular:
ind$^{D=10}($Sp$(6)=2$), but disagreement for the exceptional group
G$_2$. This means that either the mass deformation method 
\cite{porrati}, \cite{kasm} fails in this case, or that the boundary term
is a positive number. It is clearly and exciting problem to resolve
this puzzle.

\begin{center}
Table 2: $D=4$, $D=6$ and $D=10$ bulk index ind$_0^D$
and (conjectured) total Witten index ind$^D$ for the 
orthogonal, symplectic and exceptional
groups of rank~$\leq 3$. Previously unknown values are
printed in bold-face.\\
\vspace{0.5cm}
\begin{tabular}{||c|| c | c | c | c | c | c ||} \hline
Group  & rank  & ind$_0^{D=4,6}$ & ind$^{D=4,6}$ & ind$_0^{D=10}$ &
ind$^{D=10}$ \\ \hline  \hline  
SO(3)  & 1 & $1/4$  & 0  & 5/4 & 1 \\
SO(4)  & 2 & $1/16$ & 0  & 25/16 & 1 \\
SO(5)  & 2 & $ 9/64 $ & 0 & {\bf 81/64} & 1 \\
SO(6)  & 3 & $1/16$ & 0 & 21/16 & 1 \\
SO(7)  & 3 & $ 25/256 $ & 0 & {\bf 325/256} & 1 \\ \hline \hline 
Sp(2)  & 1 & $1/4$  & 0 & 5/4 & 1 \\
Sp(4)  & 2 & $ 9/64 $ & 0  & {\bf 81/64} & 1 \\
Sp(6)  & 3 & $ 51/512 $ & 0  & {\bf 1175/512} & 2 \\ \hline \hline 
G$_2$  & 2 & $ 151/864 $ & 0  & {\bf 1375/864} & {\bf 1}~? \\ \hline 
\end{tabular} 
\end{center}

In summary, we have demonstrated that Witten index calculations
for supersymmetric gauge quantum mechanics remain very subtle and
are fraught with pitfalls. In particular the available methods for
computing the boundary contribution to the index appear to be
unreliable. We furthermore found some evidence that mass deformation
methods might fail for some gauge groups as well. On the other hand,
the BRST deformation method for the bulk index appears to be fully
valid in $D=4,6,10$. It would be exciting to compute the resulting
contour integrals for a general gauge group. For $D=10$ it should be 
interesting to further check the method by numerical techniques.
Finally, it will be important to rigorously derive the proposed number
of $D=10$ vacuum states, in particular in the controversial case of
the exceptional groups.

\acknowledgements
We thank W.~Krauth and A.~Smilga for useful discussions.

\end{document}